\documentstyle[12pt]{article}
\begin{document}
\newcommand{\beq}{\begin{equation}}
\newcommand{\eeq}{\end{equation}}
\newcommand{\bea}{\begin{eqnarray}}
\newcommand{\eea}{\end{eqnarray}}
\newcommand{\ab}{\alpha\beta}
\newcommand{\cd}{\gamma\delta}
\newcommand{\ac}{\alpha\gamma}
\newcommand{\bd}{\beta\delta}
\newcommand{\mn}{\mu\nu}
\newcommand{\mpnp}{\mu'\nu'}
\newcommand{\abcd}{\alpha\beta\gamma\delta}
\newcommand{\bsigma}{\mbox{\boldmath $\sigma$}}
\newcommand{\btau}{\mbox{\boldmath $\tau$}}
\newcommand{\eg}{{\sl e.g.\ }}
\newcommand{\ie}{{\sl i.e.\ }}
\newcommand{\etc}{{\sl etc.\ }}
\newcommand{\etal}{{\sl et al}}
\newcommand{\tabh}{ \rule[-8pt]{0pt}{24pt} }
\newcommand{\cl}[1]{\begin{center} {#1} \end{center}}

\cl{\Large{{\bf Scaling and universality in binary fragmenting with
inhibition}}}
\vskip 1truecm
\cl { Robert Botet$^{ \dagger }$ and
Marek Ploszajczak$^{ \ddagger}$  }
\vskip 0.5truecm
\cl     {{$\dagger$}
Laboratoire de Physique des Solides,
B\^{a}timent 510, Universit\'{e} Paris-Sud,
\\ Centre d'Orsay, F-91405 Orsay, France
\newline           {$\ddagger$}
Grand Acc\'{e}l\'{e}rateur National d'Ions Lourds (GANIL),
BP 5027, \\
F-14021 Caen Cedex, France }
\vskip 1truecm

\begin{abstract}

We investigate a new model of binary fragmentation
with inhibition, driven
by the white noise. In a broad range of fragmentation probabilities, the
power-law spatiotemporal correlations are
found to arise due to self-organized
criticality (SOC). We find in the SOC phase
a non-trivial power spectrum of the
temporal sequence of the fragmentation events. The $1/f$~ behaviour is
recovered in the irreversible, near-equilibrium part of this phase.

\end{abstract}
\vskip \baselineskip
\vfill

\newpage

The concept of self-organized  criticality (SOC) \cite{bak}~
has attracted much attention as it has been
demonstrated that it leads naturally to power-law distributions
in space and time. The ubiquity of such
correlations in nature and, in particular, $1/f$ noise in time
\cite{flicker}~, are expected to follow from the widespread
occurence of SOC in most dissipative many-body systems.
In the fragmentation processes, the idea of universality comes
from the observation that many experimental  data
on cluster-size (mass) distribution can be fitted by a single
law $n_k \sim k^{- \tau}$~ , independently of interactions involved
or orders of scale. Values of the exponent $\tau$~
are known sufficiently accurately to exclude the possibility of grouping
in few universality classes by the value of $\tau $~.
There is here a striking resemblance
with the fragmentation - inactivation binary (FIB)
model\cite{my,my1}~, which splits the set of all sequentially
fragmenting systems into three classes. Two of them , at the
transition line and in the shattering phase, exhibit power-law
size-distribution,  but the possible values
of the exponent $\tau$~ are smaller than $2$~ in one
class (on the transition line)
and are larger than $2$~ in the  other one (the shattering phase)
\cite{info}~. Despite this
qualitative resemblance  these two classes are of fundamentally
different nature as manifested by the properties of
the cluster-size fluctuations\cite{my1,review}~.
It has been shown that the FIB model is able to reproduce quantitatively
the multifragmentation results in nuclear heavy-ion collisions
at intermediate energies\cite{my1}~.

In this Letter, we introduce the FIB automaton which is
an alternative version of the FIB model,
and show that it evolves
into a SOC state with no characteristic time or length scales
in a broad range fragmentation probabilities.
We will show also that the occurence of the anomalous
power spectra $1/f^{\beta}$~ ($1 \leq \beta < 2$)
in sequential fragmentation is
associated with the SOC phase  of the FIB automaton.
In particular, $1/f$~ spectrum is obtained
in the limit of weak fragmentation activity.

The FIB automaton is defined on a one-dimensional
lattice of linear size $N$~($i = 1, \dots, N$). The $i$th coordinate
is an integer denoting a {\it mass} of the object. To each
lattice site we assign an integer $n(i)$~ which is to represent
a {\it multiplicity of fragments} of mass $i$~.
The dynamical rules of the
FIB automaton include the perturbation mechanism and the
relaxation rules. Let us beginn by the relaxation rules.
One deals with clusters characterized by some conservative quantity,
that we shall call the {\it cluster mass}~.
The cluster of mass $N$~
is relaxing via an ordered and irreversible
sequence of steps. The first step
is either a binary fragmentation occuring
with the probability $p_F$~ or an
inactivation, with probability $1-p_F$~.
Once inactive, the cluster cannot be reactivated anymore.
The successful
fragmentation leads to two fragments,
say $(N) \longrightarrow (j)~+~
(N-j)$~, with the mass partition probability $\alpha F_{j,N-j}$~,
where $\alpha$~ is given by the normalization condition for
coefficients $F_{i,j}$~. The following step is either the fragmentation
of the cluster
$(j)$~, with probability $p_F$~, or its inactivation $(j) \longrightarrow
(j)^{*}$~, with probability $1-p_F$~. The relaxation process terminates
when all clusters are inactive.

We consider the fragmentation process which is driven by the
'white noise' ($F_{j,l} = F_{l,j} = const$). It is
assumed that the fragmentation probability $p_F$~
along the fragmentation cascade has a {\it fixed}
value between 0 and 1, independent of the size of the fragmenting
object, except for the monomers which are not
allowed to break up \cite{exp}~. Any event , either a fragmentation
or an inactivation , occurs in the time interval between
$t$~ and $t+dt$~ after the cluster has been activated
with a probability $p_F$~.
For  $p_F = const$~, the FIB process can be mapped into the directed
percolation on the Cayley tree which is a mean-field percolation.
The fragmentation is a branching process,
since the propagator of
fragmentations is independent of mass.
Each node is occupied with a
probability $1$~ and at each occupied point
at time $l~dt$~ one chooses  between three possibilities~:
fragmentation, inactivation, and 'no event' with respective probabilities
${p}^{2}$~, $(1 - p)^{2}$~ and $2p(1 - p)$~. The fragmentation
probability is~: $p_F = p^{2}/[ p^{2} + (1 - p)^{2}]$~.
This fragmentation is also analogous to the process of
self-avoiding random walk, because the previously activated
sites of this tree-like process 'repel' any subsequent reactivation.
At each fragmentation, a given
cluster is replaced by two descendants and the fragmentation
multiplicity increases by one unit. The probability $P_N({\tilde S})$~
that a FIB branching process creates exactly $M_0$~
fragments is for large $N$~:
\bea
\label{e0}
P_N[M_0 = {\tilde S}] & = & \frac{1}{2 p_F}~\frac{1}{2{\tilde S} - 1}
\left( \begin{array}{c} 2{\tilde S} \\ {\tilde S} \end{array} \right)
[ 4 p_F (1 - p_F) ]^{{\tilde S}}
\sim {\tilde S}^{-3/2} \exp(\alpha {\tilde S})
 ~~~~~\mbox{if}~~~~ p_F \leq 1/2
 \nonumber   \\
\eea
with $\alpha = \log[4p_F(1 - p_F)]$~.
For $p_F > 1/2$~, the branching tree has an infinite size ($M_0 =
\infty $~). At last, $p_F = 1/2$~ is the critical point of
the FIB branching
process as in this case the branching tree barely survives and
a fragmentation multiplicity for large $N$~ is a power law with
the exponent $3/2$~.
Moreover, the average value of the multiplicity
, which is $<M_0> = (1 - p_F)/(1 - 2p_F)$~ for $p_F < 1/2$~
becomes infinite if $p_F > 1/2$~. The value
$p_F = 1/2$~ is also a critical point of the avalanche process
for the undirected sandpile process
in the mean-field limit\cite{obukh}~. Notice however, that
the sequential fragmentation
is also critical from the point of view of the shattering transition
for any value of $p_F$~,
even though the branching tree size-distribution is modified.

The FIB automaton is perturbed by adding one mass unit to the
randomly chosen fragment $(j)$~ on the lattice
$\{ n(i),~i=1, \dots, N\}$. The cluster of mass $N$~ is
unstable and, therefore, the fragmentation avalanche is started
if the perturbation produces a cluster $N$~. Once we have
chosen the perturbation mechanism and the relaxation algorithm then
the algorithm of the temporal evolution
of the FIB automaton goes as follows~: \newline
(1) Specify an initial configuration $\{ n(i),~ i = 1, \dots ,N \}$~.
\newline
(2) Whenever $n(N) \neq 0$~, the cluster of size $N$~ breaks into
smaller fragments according
to the relaxation algorithm of FIB model until all branches of the
FIB cascade become  inactive.\newline
(3) The fragment multiplicity at all sites
is updated, increasing by the amount of produced inactive clusters.
\newline
(4) Choose a fragment $(j)$~
at random and increase its mass by one unit:
$j \rightarrow j + 1$~. Change the fragment multiplicities at sites
$j$~: ~$n(j) \rightarrow n(j) - 1$~
and $j+1$~: ~$n(j+1) \rightarrow n(j+1) + 1$~.
Return to step $2$~.    \newline
For $p_F < 1/2$~, after a transient period,
a stationary state of FIB automaton
is reached in which the normalized
average size of the cluster $\sigma(t) = (1/N) \sum_{j} j~n(j)$~
exhibits small
fluctuations around its asymptotic value~:
\bea
 \bar{\sigma}
= \lim_{T \rightarrow \infty} \frac{1}{T} \int_{0}^{T} dt
 \sigma(t) ~=~ \frac{1 - 2p_F}{2 - 2p_F}   ~~~~.
\eea
This is the 'high-viscosity' regime, where the accumulation
mechanism of the FIB automaton compensates exactly the
relaxation mechanism and the system drives alone into
a steady state with parameters independent of the lattice size.
We shall see below that the steady state of the FIB automaton in this
regime has all features of SOC state.
In the upper part of Fig. 1 we show for $p_F = 0.1$~
the dependence of the average size of the clusters on time, where
a unit time-step is one update of the whole lattice
$\{n(i),~i = 1, \dots, N\}$~.
For $p_F \geq 1/2$~, the asymptotic average cluster-size $\bar{\sigma}$~
depends explicitely
on the lattice size and equals $ \sim (\ln N)^{-1}$~ for
$p_F = 1/2$~ and $\sim 1/[(2 - 2p_F)\zeta(2p_F)] N^{1 - 2p_F}$~ for
$p_F > 1/2$~. Thus, $\bar{\sigma}$~ becomes equal zero in
the thermodynamic limit $N \rightarrow \infty$~.
This is the
'low-viscosity' regime of the FIB relaxation mechanism, where the
cluster-mass accumulation is completely destroyed.
We have found , that for any $p_F$~,
the cluster-size distribution in the
steady state of the FIB automaton exhibits a power-law behaviour.
However, in contrast to the original FIB model \cite{my,my1}~, it
does not manifest any finite-size effects even in small lattices
(see the lower part of Fig. 1). The average slope of the cluster-size
distribution in the steady state exhibits small fluctuations about its
asymptotic value $\bar{\tau} = 2 p_F$~.

Before we discuss the statistical properties of FIB automata
, let us first define
the {\it instanteneous dissipation rate} for the
avalanche $e$~:
\bea
\label{e3}
f_{(e)}(t) = \sum_{k}^{} \chi_{k}^{(e)} (t)
\eea
$f_{(e)}(t)$~ is an indicator function of unstable clusters at a
time $t$~ in the fragmentation avalanche.
$\chi_{k}^{(e)}(t)$~  in (\ref{e3})
is the characteristic function of the cluster $k$~ and
equals either $1$~, for $t \in [t_k, t^{'}_k]$~ where $t_k$~ is
the instant of formation of the cluster $k$~ and $t^{'}_k$~ the
time when it disappears, or $0$~ otherwise.
The summation goes over all clusters in the avalanche $e$~.
At a given time $t$~
, the value of the characteristic function
$f(t)$~ is then just the number of {\it active} clusters.

The lifetime distribution of fragmentation avalanches are shown
in Fig. 2~ for few selected values of $p_F$~.
One may notice
a power-law behaviour $P[T = t] \sim t^{1-{\alpha}^{'}}$~ for
$p_F < 1/2$~,  \ie in the 'high-viscosity' regime. The values
of the exponent ${\alpha}^{'}$~ for different values of $p_F$~
are summarized in Table 1.

The avalanche-size is defined as the {\it total dissipation } of the
fragmentation avalanche~:
\bea
\hat{s} = \int_{0}^{\infty} f(t) dt = \sum_{k}^{}~({t'}_k - t_k)
{}~ \sim ~ {\tilde M_0}~{\tilde T}
{}~~~~,
\eea
This is just the sum of the lifetimes
of all the clusters which have appeared in the sequence of breaks,
where ${\tilde M_0}$~  and ${\tilde T}$~ are
the average multiplicity and the total fragmentation time for a
given event respectively.
For $p_F < 1/2$~ , the value of ${\tilde M_0}$~ is a function
of $p_F$~ but does not depend on the size of the fragmenting object.
Hence, we expect that $\hat{s} \sim t^{\gamma_1}$~ with $\gamma_1 = 1$~.
This is verified by the calculated size-distribution,
which exhibits also a power-law
behaviour $P[\hat{S} = \hat{s}] \sim {\hat{s}}^{1-{\tau}^{'}}$~
with ${\tau}^{'} = {\alpha}^{'}$~.
For $p_F > 1/2$~, \ie in the 'low-viscosity' regime,
the branching tree of the fragmentation process
survives until the low mass cutoff for monomers. In this case,
the distribution of both avalanche-lifetimes and
avalanche-sizes loose their scale-invariant features and are dominated
by the cutoff scale. Consequently,
only for relatively small time and size scales, the
distributions $P[T=t]$~ and $P[\hat{S}=\hat{s}]$~
are consistent with a power-law behaviour. In this range of scales,
 ${\alpha}^{'} = {\tau}^{'} = 2$~, independent of $p_F$~.

The sequential fragmentation in FIB automaton
has properties of the SOC state for all $p_F$~ in the 'high-viscosity'
regime $0 < p_F < 1/2$~.  For each $p_F$~, the stationary
state of the FIB automaton has different features as manifested
\eg by different asymptotic values of the average cluster-size
$\bar{\sigma}$~ or the average slope of the cluster-size
distribution $\bar{\tau}$~.
For each value of the fragmentation
probability $p_F$~ ($< 1/2$), FIB model provides a different
mean-field realization of a SOC state with its particular
exponents of the power-law spatiotemporal distributions.
In this sense, it is appropriate
to speak about a SOC phase of FIB automaton for $0 < p_F < 1/2$~.
The extreme points: $p_F = 0$~ and $1/2$~ at the two
ends of the SOC phase
have a particular significance as they correspond to the
stationary limit of the fragmentation process and to the critical
point of the branching process respectively. Below, we shall show,
that they play also a special role in characterizing the anomalous
power spectra $S(f) \sim 1/f^{\beta}$~
with $1 \leq \beta \leq 2$~.

One of the main aims behind SOC studies was the search for a generic
explanation of $1/f$~ flicker noise\cite{flicker}~ whose
widespread occurence is one of the great unresolved problems
in physics. The concept of SOC state is an attempt to find
a general principle behind ubiquity of $1/f$~ noise in apparently
unrelated physical systems.
As pointed out before\cite{jens,jens1,christ}~, the spatiotemporal
scaling in the SOC state does not necessarily manifest itself in the
nontrivial features of the noise spectrum. Indeed, the power spectrum
of the cellular automaton model of Bak, Tang and Wiesenfeld\cite{bak}
is $1/f^{2}$~ as for the random walk.

Power spectrum of linearly superimposed fragmentation avalanches
can be calculated directly. For that let us define the {\it total
dissipation rate}~:
\bea
j(t) = \sum_{e}^{} f_{(e)}(t - \tau_{(e)})
\eea
where $\tau_{(e)}$~ is a random instant of time
when a fragmentation in the event
$e$~ has started and the summation goes over all events generated by the
FIB automaton. The power spectrum of $j(t)$~:
\bea
S(f) = < \left( \sum_{k}^{} \frac{\cos f{t'}_k - \cos f t_k}
{f} \right)^{2} + \left( \sum_{k}^{} \frac{\sin f {t'}_k -
\sin f t_k}{f} \right)^{2} >
\eea
is shown in Fig. 3 for few values of $p_F$~. The
function $S(f)$~ indeed shows a power law
behaviour $\sim 1/f^{\beta}$~ even in relatively
small systems $(N = 2048)$.
The values of the power law
exponent $\beta$~ for different values of $p_F$~ are
shown in Table 1.
For $p_F \geq 1/2$~, the power spectrum is those
of a random walk, \ie $1/f^{2}$~. For $p_F < 1/2$~,
\ie in the SOC phase of the FIB automaton, the power spectrum
is anomalous ($1 < \beta < 2$~) in the frequency range
$f_0 < f < f_1$~. For $f < f_0$~ , one finds a 'white noise'
 power spectrum and the low-frequency limit $f_0$~
is independent of $N$~. For $f > f_1$~,  the power spectrum
becomes $1/f^{2}$~  and  the high-frequency limit
$f_1$~ grows linearly
with $N$~. High frequencies correspond to short fragmentation
times, where one is
sensible to the fragmentation of the first cluster of size $N$~.
Since this process is unique, one expects $1/f^{2}$~ spectrum
as soon as $f \gg 1/\tau_N$~, where $\tau_N \sim p_F/(N - 1)$~
is a typical time of disappearing of the first cluster in the
fragmentation avalanche.  The exponent $\beta$~
becomes $1$~ in the limit $p_F \rightarrow 0$~, \ie
when approaching the stationary limit  of the FIB automaton.
The exponent $\beta$ of the power spectrum $S(f)$~
is related to the exponents ${\alpha}^{'}, {\tau}^{'}$~
of the lifetime- and size- distributions by the relation~:
\bea
\label{ch1}
\beta = (4 - {\tau}^{'})~\frac{2 - {\alpha}^{'}}{2 - {\tau}^{'}}~~~.
\eea
This relation has been derived by Christensen et al.
(see eqs. (36), (39) and (40) of ref.\cite{jens1}~) for sandpile
cellular automata. Numerical verification of this relation is
shown in Table 1. The validity of (\ref{ch1})
for $p_F \geq 1/2$~ is less certain as the power-law behavior
in distributions $P[T = t]$~ and $P[\hat{S} = \hat{s}]$~
is here restricted to relatively short times only.

The proposed mechanism for anomalous activity spectrum due to
critical fragmentation
is a `bulk' phenomenon associated with the `mass' fragmentation.
It is an irreversible phenomenon, associated with the dynamical
SOC phase of the fragmentation process. The coherence in the activity
spectrum over all time scales ($\beta \rightarrow 1$) is found
in the limit of dying out fragmentation activity ($p_F \rightarrow 0$),
\ie close to the stable equilibrium of the system without external
driving. The ubiquity of the $1/f$~ phenomena follows once
it has been established that the near-equilibrium dynamics of
most dissipative many-body systems is determined by the SOC state
which is an attractor to the complicated many-body dynamics.
In contrast to the explanation
of the flicker noise, based on the deterministic spring - block
models\cite{olami}~ or the lattice-gas model\cite{jensen}~
the present model has also an inherent local conservation
law, the cluster `mass' in each fragmentation vertex.

\vskip \baselineskip

\newpage

{\Large {\bf Figure captions}} \\

{\bf Fig. 1} \\
In the upper part,
the normalized average size of the cluster $\sigma$~ is plotted
as a function of time  for the FIB automaton with
$N = 512$~ and $p_F = 0.1$~. The solid line shows the asymptotic
value $\bar{\sigma} = 4/9$~ when $t \rightarrow \infty$~.
The bottom part of the figure shows
the cluster-size distribution of the FIB automaton
in the steady state  for $p_F = 0.1$~ (filled circles) and
$0.25$~ (filled triangles) in the SOC phase,
for ~$p_F = 0.5$ (open circles)  at the critical point of
the branching process
and for ~$p_F = 0.75$ (open squares) above it.
The slope of the power-law distributions $n(k)$~ is
$\tau = 2 p_F$~ , in accordance with results found earlier
in the FIB model\cite{my1}~.
Notice a total absence of the finite-size effects
even for $k = N$~. The data correspond to $10^{6}$~ fragmentation
avalanches with $N = 256$~.
 \\

{\bf Fig. 2} \\
Lifetime distributions of the fragmentation avalanches for different
values of $p_F$~. Significant deviations from the power-law
distribution for large times
are seen for $p_F \geq 0.5$~ and are due to the appearance
of 'infinite' branching trees which reach the cutoff for monomers.
The data correspond to $10^{6}$~
fragmentation avalanches with $N = 2048$~.
The same signatures are used as in the bottom part of Fig. 1.
\\

{\bf Fig. 3} \\
Power spectra corresponding to the lifetime distributions
shown in Fig. 2. The same signatures are used as in the bottom
part of Fig. 1.
\\

\vspace{2cm}
{\bf Table 1}\\
Exponents for different values of the fragmentation probability
$p_F$~. In the last column, the sum of calculated exponents $\beta$~
and ${\tau}^{'}$~ is given. $\beta + {\tau}^{'} = 4$~ for sandpile
cellular automata as verified by Christensen et al\cite{jens1}.  \\
\vskip1.5cm
\begin{tabular}{|r|c|c|c|}
\hline
$p_F$ &  $\beta$ & ${\alpha}^{'} = {\tau}^{'}$ & $\beta + {\tau}^{'} $ \\
\hline
$0.05$ & $1.0 \pm 0.1$ & $ 3.0 \pm 0.3 $  & $  4.0 \pm 0.4$\\
$0.10$&$1.16 \pm 0.05$&$2.81 \pm 0.05$  & $  3.97 \pm 0.1$\\
$0.25$&$1.50 \pm 0.03$&$2.53 \pm 0.10  $& $  4.03 \pm 0.1$\\
$0.40$&$1.91 \pm 0.05$&$1.90 \pm 0.10$  & $  3.81 \pm 0.1$\\
$0.50$&$2.00 \pm 0.02$&  ------  &  ------  \\
$0.75$&$2.00 \pm 0.05$&  ------  &  ------  \\
\hline
\end{tabular}


\begin{thebibliography}{99}
\bibitem{info} The third class, associated with the $\infty$ - cluster
phase, cannot be described in terms of the scale-invariance.
\bibitem{exp}
An absence of the size dependence in  $p_F$~ is equivalent to
the time-independence of the fragmentation probability,
which in turn follows from the criticality of the general
FIB model at the shattering transition\cite{my,my1}~.
\end{thebibliography}
\end{document}